# From Mechatronic Components to Industrial Automation Things

An IoT model for cyber-physical manufacturing systems

Kleanthis Thramboulidis, Theodoros Foradis

Electrical and Computer Engineering,
University of Patras,
Patras, Greece

*Abstract*—IoT is considered as one of the key enabling technologies for the fourth industrial revolution, that is known as Industry 4.0. In this paper, we consider the mechatronic component as the lowest level in the system composition hierarchy that tightly integrates mechanics with the electronics and software required to convert the mechanics to intelligent (smart) object offering well defined services to its environment. For this mechatronic component to be integrated in the IoT-based industrial automation environment, a software layer is required on top of it to convert its conventional interface to an IoT compliant one. This layer, that we call IoTwrapper, transforms the conventional mechatronic component to an Industrial Automation Thing (IAT). The IAT is the key element of an IoT model specifically developed in the context of this work for the manufacturing domain. The model is compared to existing IoT models and its main differences are discussed. A model-to-model transformer is presented to automatically transform the legacy mechatronic component to an IAT ready to be integrated in the IoT-based industrial automation environment. The UML4IoT profile is used in the form of a Domain Specific Modeling Language to automate this transformation. A prototype implementation of an Industrial Automation Thing using C and the Contiki operating system demonstrates the effectiveness of the proposed approach.

## I. INTRODUCTION

Based on one of the most commonly used definitions, the term Mechatronics emphasizes on the synergistic integration of the three discipline areas, i.e., mechanical engineering with electronics and intelligent computer control in the design and manufacture of products and processes [1], i.e., it emphasizes on synergy. What is not clear by this definition is the level at which this integration should be performed, i.e., at the system level, which is the traditional approach or at the subsystem or even at the mechanical unit level. The latter approach is proposed in Model Integrated Mechatronics [2] and refined with the 3+1SysML-view model [3][4]. This approach defines the Mechatronic component as the main building block that abstracts the physical world object to the software level, adding at the same time additional functionality to the one offered by the mechanical part, transforming it to a smart object. The so constructed mechatronic components are integrated with cyber components and humans to construct the industrial automation system. This approach slightly finds its road to production in the context of Industry 4.0, e.g., FESTO [5], since it greatly reduces the coupling between the system components compared to the traditional one, which considers the integration of the three disciplines at the system integration level.

A great number of communication mechanisms and middlewares are used for the integration of the constituent components of mechatronic systems in the industrial automation systems (IAS) domain. However, last years with the evolution of IoT there is a trend to exploit the benefits of this technology. IoT is aligned well with the architecture of a manufacturing enterprise and it is able to provide "vital solutions to planning, scheduling, and controlling of manufacturing systems at all levels." [6]. Several approaches consider IoT as a technology that can be utilized as integration mechanism to be used down to the sensor and actuator level of the industrial automation system. Others, consider IoT as the new logical transition from the automation and connectivity concepts that exist in the IAS domain for many years. Authors in their article with title "The Internet of Things – The future or the end of mechatronics" [7] argue that many of the smart components associated with the IoT will be essentially mechatronic in nature, and will be constructed as far as it regards their interaction with the physical world on the conventional hierarchical model. This model considers the controller of the mechatronic component in the loop with the controlled physical unit through sensors and actuators.

However, for the conventional mechatronic component to be integrated in the IoT-based industrial environment a software layer is required on top of it to convert its conventional interface to an IoT-compliant one. In this way, the adoption of the IoT as integration technology for the system transforms the conventional mechatronic component to an Industrial Automation Thing (IAT). This transformation is more likely, as authors also argue in [7], to bring significant changes to the way mechatronic, and related, systems are designed and configured. There is already an increasing complexity in the job of the industrial engineer in the task of transferring the functionality of the physical world in the software world in the level of the IAT. To this, the complexity of adding an extra layer to transform the conventional mechatronic component to an IAT is added. New protocols, languages environments and architectural paradigms should be used and successfully integrated with the already used conventional architectures and this complicates the job of industrial engineer.

Authors in [23] present UML4IoT with focus on the modeling of the Industrial Automation Thing. UML4IoT is a UML-based approach that realizes the model driven engineering paradigm to exploit IoT in the manufacturing domain. In this paper, (a) we extend the model of the IoT introduced in [23], and (b) define a model-to-model transformer to automate the construction of IATs based on the Contiki Operating system [9] and the C language. The Industrial Automation Thing is still the key artifact in this extended model for the adoption of the IoT infrastructure in the manufacturing domain. The manufacturing system is considered as a composition of cyber-physical and cyber components along with humans [13]. All these components are considered as Things, either permanent or on demand, that collaborate exploiting an IoT communication infrastructure to realize a higher level of behavior, i.e., the one of the system level. The presented approach is discussed in comparison with other approaches and mainly the IoT-A reference architecture that has been adopted by the Papyrus for IoT project [24], which is building a platform for the design of IoT systems in general. This project has many similarities with our project; both projects use UML and SysML as modeling languages, and Papyrus as tool to provide a modeling solution for IoT. Our approach focuses on manufacturing systems.

This paper focuses on the case that a high level design specification for the mechatronic component is not available. Specific annotations were defined to annotate the C source code specification of the mechatronic component so as to automatically transform the mechatronic component to an IAT. The approach is presented using as case study the Liqueur production laboratory system. The LWM2M IoT application protocol [10] running on top of the Constrained Application Protocol (CoAP) [11] is used as IoT protocol stack. The remainder of this paper is structured as follows. Section 2 positions this research against related work. In Section 3, the proposed extension to the IoT model used in the UML4IoT approach is presented and discussed in comparison with existing IoT models. In Section 4, the Contiki based Industrial Automation Thing is presented along with the case study. The model-to-model transformer for the C language and the Contiki operating system is presented in section 5 and the paper is concluded in the last section.

## II. RELATED WORK

IoT offers in the industrial domain new levels of connectivity that may lead to higher efficiency, flexibility, and interoperability among industries [14]. However, not only many definitions exist for the IoT but also several models. These models, e.g., ETSI, IETF, SENSEI, have been developed to capture the key concepts of the domain and provide the infrastructure to develop frameworks and architectures for the systems based on IoT. As authors claim in [25], IoT has been used in various application domains although there is still no clear and uniform definition and architecture about it. A detailed discussion on the IoT models can be found in [26], where the IoT-A reference model for IoT is presented and validated. Applications of IoT in Manufacturing are presented and discussed in [27] where authors present a five layer architecture for manufacturing based on IoT.

In [15] authors focus on the device nature of Thing and consider sensors, actuators and controllers as IoT devices, i.e., things. They focus on the data field structures and evaluate the benefit of using an IP smart gateway as the decentralized peripheral to integrated sensors, actuators and the controller and claim that this may improve the performance of IoT devices. We do not agree with the use of sensors and actuators as first class model elements in the high level design specification of the system. Sensors and actuators are just technology artefacts used to integrate the physical with the cyber world so they have no place in the high level design spec of the system. In our approach, the Industrial Automation Thing, which encapsulates sensors, actuators and the controller, plays the role of the Thing and represents the key construct in the IoT manufacturing environment.

MDD becomes more and more popular in the development of embedded software systems and various reports refer efficiency gains, from up to 50%, for example, in the development in the car industry [16], with high error reductions and a rapid increase of the maturity level of developed products. MDD is considered as a promising solution to address the complexity of software development in IoT [17] and improve quality characteristics of the produced software. MDD has already been recognized as the right tools to address the complexity of wireless Sensor Networks development exploiting abstraction, reuse, separation of concerns and automation [18]. Existing MDA approaches in this area as well as a systematic study and a classification of them based on a comparison framework can be found in [18]. Several works publish results that exploit the MDD paradigm in IoT based systems to improve their quality characteristics but also the ones of their development process.

Authors in [17] present FRASAD, a framework based on MDA to manage the complexity of IoT applications. They present a rule based model and a domain specific language to describe the application using as key concept the sensor node. The primary objective is to model the sensor node software. Contiki is also supported among other OSs by this framework. Authors assume that the application logic of the sensor node program is captured in a Platform Independent Model (PIM) and they have defined a Domain Specific Language (DSL) to map this PIM to the specific platform where it is indented to be executed. They construct the PIM using a set of rules they have defined to describe behavior of the sensor node programs. However, the approach focuses more on the message dissemination compared to the processing which is considered as an optional part of a sensor node. Our approach focuses on the interface of the Industrial Automation Thing; the behavior which is very complicated compared to one of a sensor node is defined using another DSML we have defined for structuring the mechatronic component [13].

In [19] authors present the software architecture of a platform developed to address issues, among which the lack of development toolkits, that limit the diffusion of IoT within industrial environments. They also describe an innovative, IoT oriented, model driven development toolkit that focuses on the seamless integration of heterogeneous industrial devices and sensors, into existing legacy systems by transforming them into web services. The proposed toolkit allows inexperienced developers to discover and compose distributed devices and services into mashups using a modeling tool. Thus the use of the MDD approach is mainly on the generation of the mashups and does not focus on the modeling of a mechatronic component as is the case of our approach. Furthermore, authors do not refer or describe the domain modeling language that they use in their MDD approach.

A very early approach to model complex IoT systems with UML and then generate RESTful interfaces from these models is presented in [20]. Authors do not define any DSML but they construct class diagrams and state charts using only primitive UML model elements. In [21] authors describe an approach to define a visual DSML for the IoT based on UML. They model the Thing, which they consider as key construct for building an IoT system, using the UML component construct and its interface using provided and required interfaces. Authors do not address the mapping of the conventional object-oriented (OO) interfaces of the Thing with the ones of the REST paradigm.

To the best of our knowledge there is no other work that focuses on the automation of the transformation process of the conventional mechatronic component by use of an MDA approach to an IoT compliant one that will transform it to an Industrial Automation Thing ready to be integrated into the IoT-based industrial environment.

III. TOWARDS AN IOT MODEL FOR MANUFACTURING

A. *The IoT-A reference model*

Authors in [28] describe the key concepts of the IoT-A reference architecture that is a result of a EU funded IoT project. These concepts and their interrelations are depicted in Fig. 1. Based on this, device is attached to entity, which is associated with resource that is accessed through service. In more detail, authors consider the entity as the 'thing' in the IoT, i.e., the main focus of interactions by humans and/or software agents. The device represents the hardware component that is either attached to an entity or it exists in its environment and monitors it. The resource is the actual software component that provides information on the entity or enables the controlling of the device. A service exposes the functionality of a device by accessing its hosted resources.

B. *The proposed IoT model*

Fig. 2 captures the high level key concepts of the proposed IoT model. Based on this the IoT is defined as a composition of Things and a processing and communication (IPV6-based) infrastructure (*CommunInfr*). Any artifact that is able to communicate with other Things using the processing and IPV6-based communication infrastructure (*Proc& ComnInfr*) is considered as Thing. The objective of this communication is to collaborate with these Things in order to achieve higher level of behavior compared to the one offered by each one of the collaborating Things. Collaborating Things form a new *Thing* of type *SystemAsThing* that represents a system of Things.

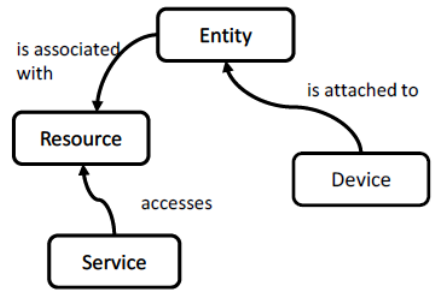

Fig. 1. Key concepts and interactions in the IoT-A model [28].

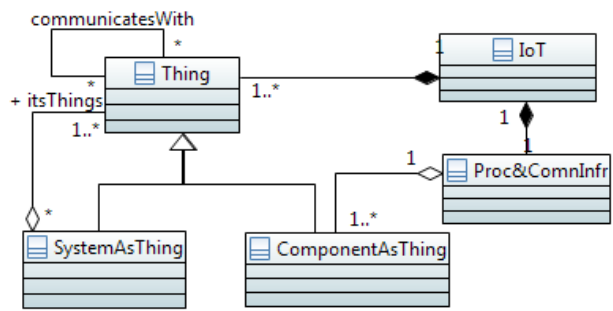

Fig. 2. High level key concepts in IoT.

A Thing may be either a system of Things (*SystemAsThing*) or a component (*ComponentAsThing*). A *ComponentAsThing* is a Thing that does not utilize IoT for the integration of its constituent components. This type of Thing is used to represent in the IoT world conventional systems or components that have been transformed to Things (*ComponentAsThing*) by adding on top of their conventional interface an IoT-based one. The smallest Thing of this type is a sensor or actuator. The *Proc& ComnInfr* is a composition of: (a) processing nodes and (b) communication devices, i.e., gateways, bridges, switches, routers, etc. It is considered as a composition of Things (*ComponentAsThing*) assuming that these devices are IoT enabled; this will be the case in the near future. The Cloud is part of the *Proc&ComnInfr*.

C. *The model of Thing in manufacturing*

Fig. 3 presents the proposed model for the Thing. Based on this, a Thing is either real, cyber or virtual. A *RealThing* is either permanent or on demand Thing. A permanent Thing is a composition of a cyber-physical object (*CpObject*) and a cyber IoT enabler (*CyberIotEnabler*). An *OnDemandThing* is an aggregation of a *PhysicalObject* and a cyber-physical IoT enabler (*CpIotEnabler*). As cyber-physical IoT enabler we model any device such as laptop, tablet, mobile phone, wearable, RFID reader, that provides an IoT like interface and is able to interact with a physical object. As physical object we mean a human, an inanimate object or even animal with an embedded or attached tag. A human interacts with an app, i.e., application specific IoTwrapper and is temporarily transformed to a Thing. Physical objects of type animals or inanimate objects with an embedded or attached tag interact with an RFID reader with IoT like interface for the same reason. A cyber-physical object is any physical object that: (a) implements some kind of functionality, i.e., material and/or energy transformations and (b) has been transformed to a smart object by appending on it information processing functionality. A mechatronic component is an example of cyber-physical object. A cyber IoT enabler integrated with a virtual objects and deployed on a processing unit constructs a virtual Thing (*virtualThing*). The IoTwrapper of the UML4IoT approach [23] is an example of cyber IoT enabler. A *cyberThing* is defined as a composition of one *CpIotEnabler* and one-to-many *cyberObjects*. This allows the developer to optionally group cyber objects of the system design model and map these to one *cyberThing*.

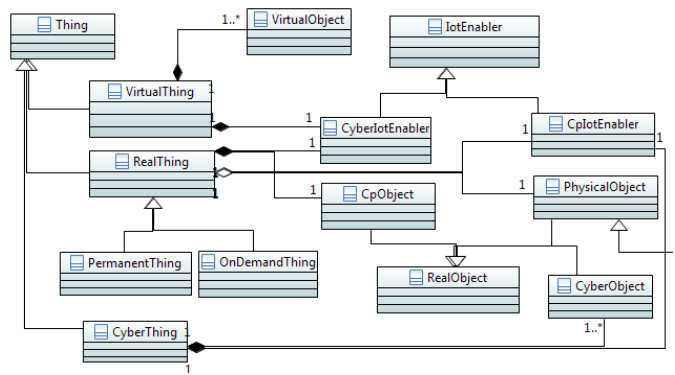

Fig. 3. The model of Thing in IoT.

A Thing may expose to its environment: (a) part of its structure in terms of properties and/or (b) part of its processing or storage functionalities. These functionalities are exposed as services. Services are discriminated to: (a) IoT infrastructure services (*IotService*), (b) application domain services and (c) application specific services. All types of services may be managed (activated, configured, updated, etc.) through the IoT communication infrastructure. Communication infrastructure services may be considered similar to the Industrial Automation Thing services with the remark that Industrial Automation Things perform material, energy and information processing while communication infrastructure Things perform only energy and information processing. For example, device management services, defined by the OMA LWM2M [10] are *IotServices*. Native services are services that would be defined for a specific application domain, e.g., home automation, manufacturing, or system specific services, such as the generateLiqueurTypeA service of the Liqueur Plant laboratory production system. Device management is implemented by the management interface of the LWM2M. On the other side, Thing management is domain or application specific and should be implemented by specific cyber components on top of the device management and service interface, e.g., the one of LWM2M.

The IoT processing and communication infrastructure (*Proc&ComnInfr*) should provide an environment for a service-based collaboration of Things. Each *Thing* implements functionalities offered to the environment as services with negotiated QoS that should be discovered and exploited by other Things. It may also utilize services of other Things to realize its behavior. The cyber components of the manufacturing system model are deployed during the deployment time on Things of type *ComponentAsThing* (see Fig. 2). UML/SysML design models of the system are marked with the IoT model elements using the UML4IoT profile to automatically transform the system to an IoT based.

### D. Discussion on the proposed IoT model

Our definition for the Thing is different from a widely accepted one and described in [28], where the authors define the thing and its relations to devices, resources and services. Device, Resource, Service and Thing are also model elements in the IoT model presented in [29]. In our approach the *ComponentAsThing* encapsulates and hides the means by which sensing of its physical part as well as actuation are realized, since this is an implementation issue. Based on this, we do not consider sensors and actuators first class model elements in our model and we do not capture these artifacts in the design model of the system. Thus, actors of the OO approach or terminator of the SA approach are modeled either as Industrial Automation Things or as human Things.

This definition of Thing satisfies the requirement set in [30] according to which all interactions among the system constituent components, which maybe humans, machines and products should be performed under the same umbrella. This allows the developers to focus on the systems functionality and do not worry about interactions increasing the productivity.

In the design model of the system we do not capture resources. A resource is a technology artifact used: (a) to represent the exposed properties and services of a Thing, and (b) to access these through a well defined set of operations to achieve low coupling between collaborating Things. The LWM2M defines a set of such operations, i.e., READ, EXECUTE, WRITE, etc., implemented on top of the http operations.

The RESTful as well as the SOAP paradigms can be utilized for accessing the services offered by Things. Thus, we adopt a different meaning for service from the widely accepted and described, e.g., in [10], where access to resources from the outside world finally happens through *services*. Author in [26] which is a result of the IoT-A EU project, consider the service as an entity that accesses a Resource which is associated with an Entity that has attached a Device. It should also be noted that while a resource is defined in [26] as the core software component that represents an entity in the digital world, a Device is attached to a Resource. We do not adopt this model because it is technology driven. Our model focuses on the system modeling level and its objective is to offer a platform independent modeling of the target system. In our model, a Thing has structural (attributes) and behavioral properties (functions/methods). Those properties that are accessible from its environment are represented as resources. The RESTful paradigm is adopted for accessing the resources. In this context and in order to exploit the benefits of IoT, the networking entities of the IoT *Proc&ComnInfr* are also considered as Things (IoT-Thing *NetworkingThing*) that provide their own set of information processing services required to establish the communication infrastructure of the IoT.

Plant processes as well as other functionalities of the plant are assigned to cyber objects of the system's design model. These cyber objects may be marked as *cyberThings*. Alternatively cyber objects may be deployed on other *cyberThings* or on Things of the *Proc&ComunInfr*. In both cases the corresponding services are mapped to resources of the corresponding Thing.

Our approach differs from the one of the ebbits platform [14] that identifies the following four layers which consider required to bind the physical world with software services: (a) physical-world layer, where devices, sensors and physical-objects are captured, (b) the IoT layer, (c) the internet-of-services layer and (d) the business system mediation and product life cycle layer. The Thing in our approach encapsulates the real-world object, its information processing unit and its smart services. Our modeling of IoT for manufacturing is not compliant with the architecture presented in [25] where a layered Architecture composed of three layers, i.e., IoT, Cloud Computing based Internet of services and Application, is adopted for the CCIoT-CMfg system that authors propose. Based on our modeling, virtual Things as well as Thing (or Things) that represent the plant may exist in the cloud and have as functionality behavior that belongs to what is known as application layer.

Authors in [31] use the term virtual object to refer to the software entity that acts as a proxy of the real-world object. In our model the meaning of the virtual Thing is completely different. We use the term software representative (SR) to refer to what authors in [31] call virtual object. Authors also capture sensors and actuators as first class model elements in their IoT metamodel.

## IV. A CONTIKI BASED INDUSTRIAL AUTOMATION THING

### A. The liqueur production laboratory system

The myLiqueur production mechatronic system, used as case study in this work, is composed of the following mechatronic components: smartSilo1, smartSilo2, smartSilo3, smartSilo4 and smartPipe. The system is based on the case study initially used in [12] and then extended in [13] to be compliant with the mechatronic component concept. The smartSilo mechatronic components are reserved in couples for the production of specific types of liqueurs. SmartSilos 1 and 4 form one couple.; smartSilos 2 and 3 the other. A mechatronic component has a well defined interface through which exposes its behavior to be used by the liqueur production processes. This interface exposes the functionalities offered by the silo such as fill, empty, mix and heat. Using the common pipe for the liquid transfer among the silos is not allowed. Moreover, mixing the liquid in two silos at the same time is not permitted due to a constraint in power consumption. Implementation issues regarding the physical silo are encapsulated and hidden from the mechatronic component's environment.

The intention is to integrate the components of this conventional mechatronic system using IoT and gain from the low coupling that this technology introduces among the interacting components. The use of the IoT will also enable the system to exploit the benefits of this technology regarding the user interaction by allowing end users to produce custom types of liqueur. The end user would be able to define, through an app (*myLiqueurApp*), the production parameters of the desired type of liqueur, as shown in Fig. 4.

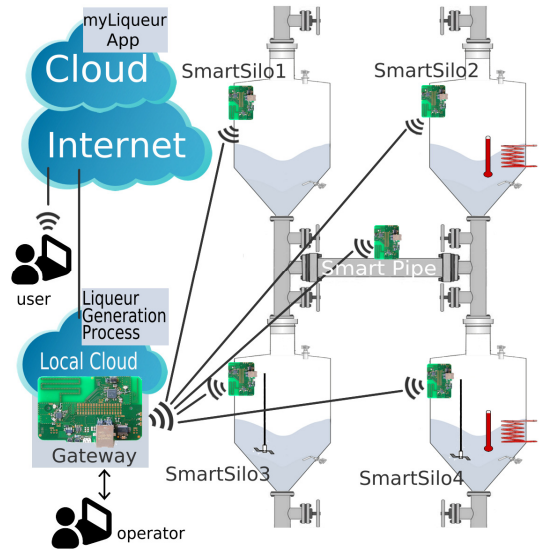

Fig. 4. The Liqueur production system used as case study

### B. The Mechatronic Component

The legacy smartSilo Mechatronic components is composed of the physical silo (physical part), a processing, storage and communication unit and the low level control software (cyber part) required for the smartSilo to provide a higher level of abstraction functionality compared to the one provided by the physical silo. As shown in figure 5, which presents the high level architecture of the mechatronic component, the software part is composed of two main parts. The first part is the software representation of the physical object, i.e., the mechanical unit, into the software domain. This part does not add any extra functionality, it only encapsulates the details of the integration of the physical world with the cyber world. On top of this, another part transforms the physical object to a smart one adding extra functionality [32]. This part encapsulates the low level control of the physical object required to transform the physical world object into a smart cyber-physical component that provides its functionality through a well defined interface.

We use the *Interface* construct of UML to specify the interface of a mechatronic component. The *Interface* is used in UML to declare a set of public features and obligations that together constitute a coherent service [22]. In this sense an

*Interface* specifies a contract that any instance of the mechatronic component shall fulfill. The UML class diagram of Fig. 6 presents the interface of the smartSilo Mechatronic component in terms of provided and required interfaces. The SmartSiloUsageIf represents the provided interface while the SmartSiloUserIf represents the required one. In the required interface we show how to model the interaction between SmartSilo and its client with the Signal and Reception constructs of UML in order to represent the possibly asynchronous nature of this interaction. Thus, the heatingCompleted and mixingCompleted signals sent by the SmartSilo will trigger an asynchronous without a reply reaction to the SmartSilo client, e.g., the type A liqueur generation process, through the corresponding Receptions captured in the SmartSiloUserIf.

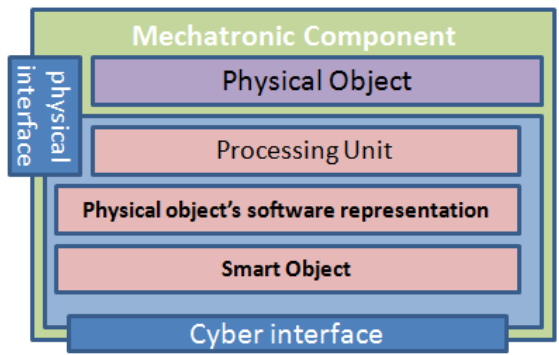

Fig. 5. The architecture of the mechatronic component

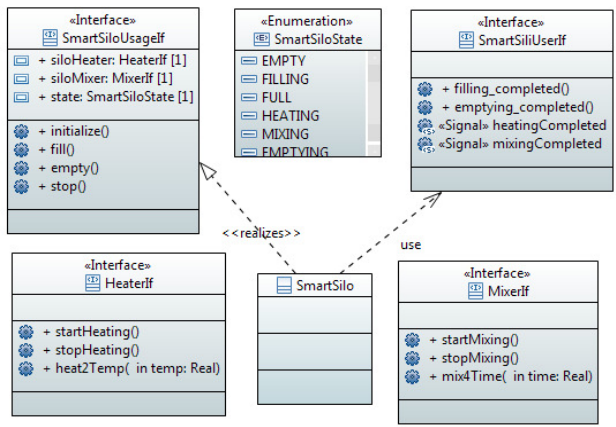

Fig. 6. The cyber interface of the SmartSilo mechatronic component

### C. Towards a Contiki-based Industrial Automation Thing

The 6LoWPAN IoT gateway of Weptech electronic Gmbh running the Contiki operating system is used to host the controller of the smartSilo mechatronic component. The 6LoWPAN IoT gateway, which is based on an ARM®Cortex® - M3 SoC with 512kB Flash and 32kB RAM, functions as a border router in a 6LoWPAN network. It connects a wireless IPv6 network, over an 802.15.4 compliant radio interface in the 2.4GHz band, to the Internet via a 10BASE-T Ethernet interface.

Contiki is a lightweight operating system ported to various microcontroller architectures on resource constraint devices [9]. It was selected mainly for its event-driven kernel that guaranties fast response times to events and to its ability for dynamic loading and replacement of individual programs and services that leads to very flexible Mechatronic components whose behavior may be modified during run-time. The interfacing of the cyber part with the physical one, i.e., the physical object's software representation, has been developed using the event-driven handling mechanism of Contiki to get a better response time compared to the traditional scan cycle approach mainly used in industry. Sensor signals generate interrupts which are handled by Contiki to transform these to asynchronous software events which are broadcasted and captured by the corresponding event handling routine that captures the response of the controller to the corresponding sensor signal. Thus the high level sensor signal is transformed to the *highLevelReached* asynchronous event that is handled by the corresponding event handling routine, which is responsible to implement the sensor data handling algorithm, that among others sends a *close* signal to the *inValve* and activates the sending of a *fillingCompleted* event to the client of the mechatronic component. The response of the system from the time that the sensor generates the signal to the time that the signal arrives to the *inValve* actuator has an average value of 39.20 μs. Fig. 7 presents a part of the object-based C implementation of the smart silo cyber part that is related with the interface of the component with its environment. This implementation is for the case that the required interface will be modeled by callback functions instead of signals and receptions. It is evident that both alternatives, i.e., signals or callback functions, imply a tight coupling among the smartSilo and the components that use its behavior, in the sense that these interfaces have to be known in advance for the development of the component's clients.

```
struct silo{
  enum silo_state state;
  struct level_sensor *high_level_sensor,*low_level_sensor;
  struct valve *in_valve,*out_valve;
  ...
  int32_t target_temperature;
  // provided If
  void (*initialize)(void);
  void (*fill)(void);
  void (*empty)(void);
  void (*stop)(void);
  void (*heat)(void);
  void (*mix)(void);
  // required If
  void (*filling_completed)(void);
  void (*emptying_completed)(void);
```

Fig. 7 Part of the C object-based implementation of the cyber part of the mechatronic component.

To automate the process of generating the IAT a transformer is required to transform the properly annotated with the DSML conventional mechatronic component. One approach is to mark the UML design specification of the mechatronic component with the stereotypes defined by the UML4IoT profile. If a UML design is not available then the source code of the cyber part of the mechatronic component is properly annotated with specific annotations that have been defined based on the UML profile. An example of annotated code with Java-like annotations is given in Fig. 8.

In Fig. 9 the Contiki-based silo industrial automation thing developed with the proposed approach is shown. A hardware simulator for the silo is used while as processing unit the Weptech embedded board is used in the current implementation. The Raspberry Pi and the XDK of Robert Bosch are alternative supported platforms.

*D. The interfaces of the Industrial Automation Thing*

Adopting the OMA LWM2M application protocol the interface of the Industrial Automation Thing is well defined and independent of the behavior that is implemented by the component. This interface is defined using UML provided and required interfaces as shown in Fig. 10. Based on this figure the IAT has three provided interfaces and three required that are independent of the nature of the component. This feature combined with the ability of dynamically loading and replacing of individual services that is supported by Contiki results to a completely flexible component regarding its behavior. New or replaced behavior can be activated by the same well defined REST interface of the Industrial Automation Thing. A comparison with Fig. 6 that captures the conventional mechatronic component interface points out the flexibility of the AIT compared to the conventional one. Through this REST interface, resources may be created and used on demand based on requirements assuming that the physical part supports the requested new behavior. Resources, Resource Instances, Objects, Object Instances which are exposed by the IAT as well as with their attributes, are accessed by the clients of the IAT through the device management and service enablement interface (*DM&SE If*).

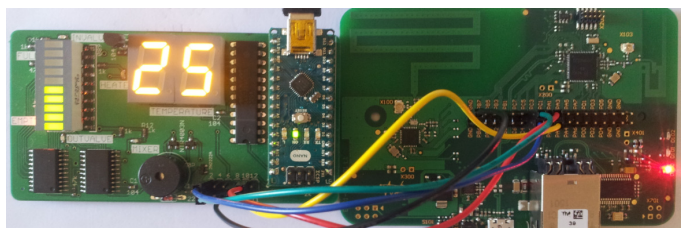

Fig. 8 Part of the C object-based implementation of the cyber part of the mechatronic component annotated with the UML4IoT java-like annotations.

Fig. 9. The silo Industrial Automation Thing based on a hardware silo simulator.

## V. A MODEL-TO-MODEL TRANSFORMER TO AUTOMATE THE GENERATION OF IAT

For the development of the source code transformer a set of transformations rules were defined to automate the generation of the Industrial Automation Thing using as input the conventional mechatronic component. The transformer transforms the annotated C source code to a LWM2M compliant Industrial Automation Thing. The following rules have been defined for the transformer. These rules apply for each object of the lwm2m client.

***Rule 1:*** Create wrapper functions for annotated behaviors.
*For each function with the BehaviorResource annotation create a wrapper function with input parameters:*
lwm2m_context_t *ctx, const uint8_t *arg, size_t argsize, uint8_t *outbuf, size_t outsize
*E.g.,* Source: `static int fill(void);`
Target: `static int fill(lwm2m_context_t *ctx, const uint8_t *arg, size_t argsize, uint8_t *outbuf, size_t outsize);`

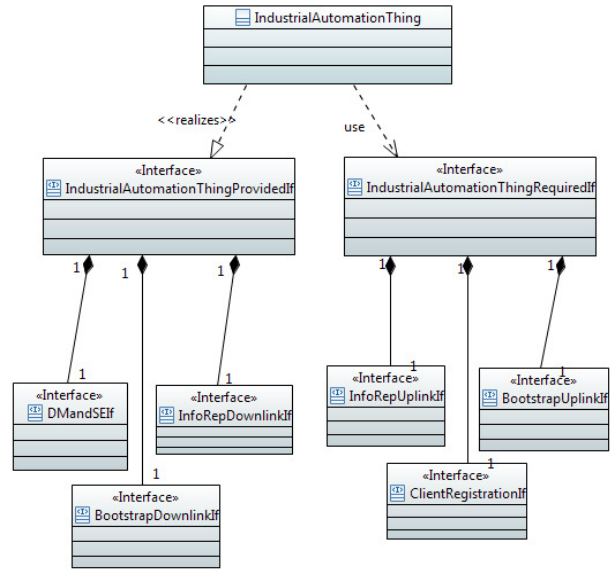

Fig. 10. Industrial Automation Thing Interfaces (provided and required).

***Rule 2:*** Create getters/setters functions for each property annotated with the *PrimitiveRes* annotation.
*For each attribute annotated with the PrimitiveRes annotation create the corresponding read and write function depending on the applied operations on the attribute defined in the annotation.*
E.g., for the silo_state property
```
static int get_silo_state(lwm2m_context_t *ctx, uint8_t *outbuf, size_t outsize) {
  char *value;
  value = get_silo_state_inString(silo->state);
  return ctx->writer->write_string(ctx, outbuf, outsize, value, strlen(value));}
```

***Rule 3:*** Construct the Resource model.
*For each annotated attribute or function create an entry using the LWM2M_RESOURCE_CALLBACK macro of the lwm2m implementation which is integrated into the Contiki OS.* E.g., for the silo_state
`LWM2M_RESOURCE_CALLBACK(0,{get_silo_state,NULL, NULL}),`
a) Append this entry to a list of resources, i.e., *silo_resources*, using the LWM2M_RESOURCES macro.

b) Create the corresponding object instance using the LWM2M_INSTANCE macro and register the resources. E.g., LWM2M_INSTANCE(0, silo_resources)

c) Append it to the list of silo instances using the LWM2M_INSTANCES.
E.g., `LWM2M_INSTANCES(silo_instances, …. );`

d) Create the Object and registers its instances using the LWM2M_OBJECT macro. E.g.,
`LWM2M_OBJECT(silo_obj, 1663, silo_instances);`

*Rule 4:* Modify setter functions.
*For each attribute annotated with the ObservableResource annotation, modify its setter function (set_<attribute name>())by appending a call to the `lwm2m_object_notify_observers` function.*
Assumption: For each observable attribute a setter function exists. E..g.,
```
Source: void set_filling_completed(){
   silo->filling_completed = 1;}
Target: void set_filling_completed(){
   silo->filling_completed = 1;
   lwm2m_object_notify_observers(&silo_obj, "/0/7");
}
```

*Rule 5:* Generate and handle the initialize function for the object.
*5.1 Generate an initialize function to initialize the legacy object and register it to lwm2m by a call to the `lwm2m_engine_register_object` function.*
The legacy initialize function of the object should be properly annotated. E.g.,
```
void ipso_ silo_init(void) {
   silo_init(); // legacy object initialization function
   lwm2m_engine_register_object(&silo_obj); }
```
*5.2 Append the initialize function prototype to the `ipso-objects.h` file. E*.g.
`void ipso_silo_init(void);`
*5.3 Append a call statement to the initialize function of each object to the `ipso_objects_init()` function body of the `ipso-objects.c` file. E.g.,*
```
void ipso_objects_init(void) {
     ipso_silo_init();
```

## VI. CONCLUSION

In this paper, we consider the tight integration of the physical world with the cyber one at the mechatronic component level. A mechatronic component offers its functionality through well defined mechanical, electrical and software interfaces. In this sense the industrial automation system is a composition of mechatronic components along with cyber components and humans. IoT is adopted for the integration of these components to exploit the benefit of this technology and UML4IoT is utilized to automatically transform the conventional mechatronic component into an IoT compliant cyber-physical one, i.e., to an Industrial Automation Thing. The IoT model used in the UML4IoT approach is extended towards a complete IoT model for the manufacturing domain. The transformation rules required for the development of the model-to-model transformer have been developed and validated through a prototype implementation of the liqueur production laboratory system. The prototype implantation of the silo industrial automation thing based on a Contiki enabled embedded board and the C language is used to demonstrate the applicability and the effectiveness of the proposed approach.